\documentclass[prd,twocolumn,showpacs,amsmath,amssymb,nofootinbib]{revtex4-1}
\usepackage{overpic,graphicx}
\usepackage{epsfig}
\usepackage{bm}
\usepackage{dcolumn}
\usepackage{bm}
\usepackage{rotating}
\usepackage{subfigure}
\usepackage{color}
\usepackage{xspace}
\usepackage{verbatim}
\begin{document}
\title{\bf \boldmath
Prospects of CKM elements $|V_{cs}|$ and decay constant $f_{D_{s}^+}$ in  $D_s^+\to\mu^+\nu_\mu$ decay at STCF
}

\author{
	\begin{small}
		\begin{center}
			\boldmath Jiajun Liu$^{1}$,  Xiaodong Shi$^{2,3}$, Huijing Li$^{4,5}$,
			Xiaorong Zhou$^{2,3}$, Bo Zheng$^{1}$\
			\\
			\vspace{0.2cm} {\it
				$^{1}$ University of South China, Hengyang 421000, People's Republic of China \\
				$^{2}$ State Key Laboratory of Particle Detection and Electronics, Hefei 230026, People's Republic of China\\
				$^{3}$ University of Science and Technology of China, Hefei 230026, People's Republic of China\\
				$^{4}$ Henan Normal University, Xinxiang 453007, People's Republic of China \\ %
				$^{5}$National Demonstration Center for Experimental Physics Education (Henan Normal University), Xinxiang 453007, China \\
		}\end{center}
         \vspace{0.4cm}
    \end{small}
}

\begin{abstract}
We report a feasibility study of pure leptonic decay $D_s^+\to\mu^+\nu_\mu$ by using a fast
simulation software package at STCF.
With an expected luminosity of $1~\mathrm{ab}^{-1}$ collected at STCF at a center-of-mass
energy of 4.009 GeV, the statistical sensitivity of the branching fraction is determined
to be 0.3\%. Combining this result with the $c\rightarrow s$ quark mixing matrix element
$|V_{cs}|$ determined from the current global Standard Model fit, the statistical sensitivity of $D_s^+$
decay constant, $f_{D_s^+}$, is estimated to be 0.2\%. Alternatively, combining the current
results of $f_{D_s^+}$ calculated by lattice QCD, the statistical sensitivity of $|V_{cs}|$
is determined to be 0.2\%, which helps probe possible new physics beyond.
The unprecedented precision to be achieved at STCF will provide a precise calibration of
QCD and rigorous test of Standard Model.
\end{abstract}

\maketitle

\section{Introduction}

The proposed Super Tau-Charm Facility~(STCF)~\cite{stcf} in China is a symmetric electron-positron collider designed
to provide $e^{+}e^{-}$ annihilation at center-of-mass~(c.m.) energies $\sqrt{s}$ from 2.0 to 7.0 GeV. The
peak luminosity is expected to be $0.5\times10^{35}$~cm$^{-2}$s$^{-1}$ at $\sqrt{s}=4.0$~GeV
and it will accumulate an integrated luminosity $(\mathcal{L})$ of more than 1~ab$^{-1}$ data each year.
By operating at $\sqrt{s}=4.009$~GeV, the STCF will produce $2.0\times10^{8}$ $D_{s}^{+}D_{s}^{-}$ with one year's data taking,
which can be used to study the purely leptonic, semileptonic and hadronic decays of $D_{s}^{+}$ with an unprecedent precision.

Among these, the purely leptonic decay $D^+_s\to \ell^+\nu_\ell$~($\ell=e$, $\mu$ or $\tau$)
offers a unique window
into both strong and weak effects in the charm sector.
In the Standard Model~(SM), the partial width of the decay $D^+_s\to \ell^+\nu_\ell$
can be written as~\cite{Silverman:1988gc}
\begin{equation}
	\Gamma_{D^+_{s}\to\ell^+\nu_\ell}=\frac{G_F^2}{8\pi}|V_{cs}|^2
	f^2_{D^+_{s}}
	m_\ell^2 m_{D^+_{s}} \left (1-\frac{m_\ell^2}{m_{D^+_{s}}^2} \right )^2,
	\label{eq1}
\end{equation}
where $G_F$ is the Fermi coupling constant, $|V_{cs}|$ is the $c\to s$
Cabibbo-Kobayashi-Maskawa~(CKM) matrix element,
$f_{D^+_{s}}$ is the $D^+_{s}$ decay constant that parameterizes the effect of the strong interaction,
$m_\ell$ and $m_{D^+_{s}}$ are the masses of lepton and $D^+_{s}$, respectively.
The determination of $\Gamma_{D^+_{s}\to\ell^+\nu_\ell}$ can directly measure the product value of  $f_{D_{s}^{+}}|V_{cs}|$ since
all other variables are known with high precision~\cite{ParticleDataGroup:2020ssz}.
One can either extract $|V_{cs}|$ by using the predicted value of $f_{D^+_{s}}$ from lattice QCD~(LQCD),
or obtain $f_{D^+_{s}}$ by using the averaged experimental value of $|V_{cs}|$.

Precise measurements of $f_{D_{s}^{+}}$~\cite{Bazavov:2017lyh,Carrasco:2014poa,Yang:2014sea} and
$|V_{cs}|$ are essential to probe new physics beyond the SM.
Currently, the averaged $f_{D_{s}^{+}}$ from various experiments indicates a $1.5\sigma$~\cite{HFLAV:2019otj}
difference from LQCD calculation~\cite{Bazavov:2017lyh},
while the latter gives a negligible uncertainty comparing to the former.
Besides, there are two standard deviations for the $|V_{cs}|$ extracted in
$D_{s}^+\to l^+\nu_{l}$~\cite{HFLAV:2019otj} and $D\to K l\nu_{l}$~\cite{HFLAV:2019otj}, which
challenges the universality for the CKM elements.
The up-to-date results of $|V_{cs}|$ and $f_{D_{s}^{+}}$ are still limited by statistics uncertainty in the measurment
of $D^+_{s}\to\ell^+\nu_\ell$~\cite{Ablikim:2021hex}. 
Future precise measurement of $D_s^+\to\mu^+\nu_\mu$ is
critical to calibrate various theoretical calculations of $f_{D^+_{s}}$ and test the unitarity of the CKM matrix.

The SM predicts the ratio of decay widths for $D^+_s\to \tau^+\nu_\tau$  and $D^+_s\to \mu^+\nu_\mu$ to be 9.75,
with negligible uncertainty.
The lepton flavor universality~(LFU) could be violated with some new physics mechanisms,
such as a two-Higgs-doublet model with the mediation of charged Higgs bosons~\cite{Fajfer:2015ixa,Branco:2011iw}
or a Seesaw mechanism due to lepton mixing with Majorana neutrinos~\cite{Branco:2011zb}.
Using the most recent experimental results, the ratio $\Gamma_{D^+_s\to \tau^+\nu_\tau}/ \Gamma_{D^+_s\to \mu^+\nu_\mu}$ is obtained to be $9.98\pm0.52$~\cite{ParticleDataGroup:2020ssz}, which is
consistent with the SM prediction within uncertainty.
However, high precise measurement of $D^+_s\to\ell^+\nu_\ell$ decay is desirable to test LFU and
other physics mechanisms beyond the SM.

In this paper, we present a feasibility study of $D_s^+\to\mu^+\nu_\mu$ decay and estimate the sensitivity of various parameters at
STCF~\cite{stcf}, where $D^+_s$ is from $e^{+}e^{-}\to D_{s}^{+}D_{s}^{-}$ at $\sqrt{s}=4.01$~GeV
with a production cross section of $\sigma_{4.01}\approx 0.2$~nb.
Though the production cross section of $D_{s}^{+}D_{s}^{-}$ is
higher, {\it e.g.} $D_{s}^{+}D_{s}^{-}$ produced via $e^{+}e^{-}\to D_{s}^{+}D_{s}^{*-}+c.c$ with
cross section $\sigma_{4.18} \approx 0.9$~nb at $\sqrt{s}=4.18$~GeV,
the pair production of $D_{s}^{+}D_{s}^{-}$ without additional particles at $4.009$ GeV helps to
reconstruct signal with better purity and free of additional systematic uncertainties coming from
$\gamma$ or $\pi^{0}$ reconstruction in $D_{s}^{*-}$ decays.

This paper is organized as follows.
In Sec.~\ref{sec:detector}, the detector concept for STCF is introduced as well as the Monte Carlo~(MC) samples used for this study.
Section~\ref{sec:ST selection} is analysis of $D_s$ candidates.
Section~\ref{sec:OPTIMIZATION} is optimization of detector response, and Sec.~\ref{results} is the results and discussion.

\section{STCF DETECTOR AND MC SIMULATION}
\label{sec:detector}
The STCF detector in design is a general purpose detector for $e^{+}e^{-}$ collider.
It includes a tracking system composed of inner and outer trackers,
a particle identification~(PID) system with $3\sigma$ charged $K/\pi$ separation
power up to 2~GeV/$c$, an electromagnetic calorimeter~(EMC) with an excellent
energy resolution and a good position resolution,
a super conducting solenoid and a muon detector~(MUD) that provides good $\pi/\mu$ separation.
The detailed conceptual design for each sub-detector can be found in Ref.~\cite{fastsimu}.

Currently, the STCF detector and the corresponding offline software system are in the research and development~\cite{Li:2021sjr}.
A fast simulation software at STCF is therefore developed to access the physics reaches~\cite{fastsimu},
which takes the most common event generator as input to perform a realistic simulation. It
incorporates the effects from tracking efficiency of charged particles and their momentum
resolution, the efficiency of PID,  the detection efficiency of photon and its energy and
position resolution, as well as kinematic fits.
The fast simulation also provides flexibly interface for adjusting performance of each
sub-system, which can be used to optimize the detector design according to physical requirements.
The process $D^+_s\to \mu^+\nu_\mu$ analysed here also serves as a benchmark process for the
optimization of detector response, {\it e.g.} tracking efficiency,
$\pi/\mu$ separation.

A pseudo-data sample, corresponding to an integrated luminosity of 0.1~ab$^{-1}$, is produced at
$\sqrt{s}=4.009$~GeV, which includes all open charm processes, initial state radiation~(ISR)
production of the $\psi(3770)$, $\psi(3686)$ and $J/\psi$,
and $q\bar{q}\,(q=u,d,s)$ continuum processes, along with Bhabha scattering,
$\mu^+\mu^-$, $\tau^+\tau^-$ and $\gamma\gamma$ events.
The open charm processes are generated using {\sc conexc}~\cite{Ping:2013jka}.
The effects of ISR~\cite{Kuraev:1985hb} and
final state radiation~(FSR)~\cite{Barberio:1993qi} are considered.
The decay modes with known branching fraction~(BF) are generated using {\sc evtgen}~\cite{Lange:2001uf}
and the other modes are
generated using {\sc lundcharm}~\cite{Chen:2000tv}.
The passage of the particles through the detector
is simulated by the fast simulation software~\cite{fastsimu}.

\section{Analysis of \boldmath{$D_s$ }  candidates}
\label{sec:ST selection}
A double-tag technique is employed to measure the absolute BF of signal process
$D^+_s\to \mu^+\nu_\mu$.
In an event where a $D^-_{s}$ meson (called the single-tag (ST) $D^-_{s}$ meson) is fully reconstructed,
the presence of a $D^+_{s}$ meson is guaranteed. In the systems recoiling against the ST $D^-_{s}$ mesons,
we can select the leptonic decays of $D^+_s\to\mu^+\nu_\mu$~(called the double-tag (DT) events).

In $e^{+}e^{-}$ collision at $\sqrt{s}=4.009$ GeV, $D^+_{s}$ mesons are produced from the process
$e^{+}e^{-} \rightarrow D^+_s D^-_s$.  Using this threshold production characteristic, we can measure
the absolute BF for $D^+_{s}$ decays with a DT method.
In this analysis, the ST $D^-_s$ mesons are reconstructed from 14 hadronic decay modes,
$D^-_s\to K^+K^-\pi^-$, $K^+K^-\pi^-\pi^0$, $\pi^+\pi^-\pi^-$,  $K^0_SK^-$,\\  $K^0_SK^-\pi^0$,  $K^-\pi^+\pi^-$,  $K^0_SK^0_S\pi^-$,  $K^0_SK^+\pi^-\pi^-$,  $K^0_SK^-\pi^+\pi^-$,   $\eta_{\gamma\gamma}\pi^-$,    $\eta_{\pi^0\pi^+\pi^-}\pi^-$,  $\eta^\prime_{\eta_{\gamma\gamma}\pi^+\pi^-}\pi^-$,  $\eta^\prime_{\gamma\rho^0}\pi^-$, and $\eta_{\gamma\gamma}\rho^-$,\\
where the subscripts of $\eta^{(\prime)}$ represent the decay modes used to reconstruct $\eta^{(\prime)}$.     Throughout this paper, the charge conjugation is always implied.

Candidate charged tracks are selected when they pass the vertex and acceptance requirements
in fast simulation.
The $K^0_S$ candidates are reconstructed from pairs of oppositely charged tracks, which satisfy a vertex-constrained fit
to a common point. The two charged tracks with minimum $\chi^{2}$ of vertex fit
are assumed to be pions produced from $K^0_S$. The $K^0_S$ is required to have an invariant mass in range $0.485<M_{\pi^{+}\pi^{-}}<0.512$~GeV/$c^{2}$. Furthermore, the decay length of the reconstructed $K^0_S$ is required
to be larger than 2$\sigma$ of the vertex resolution away from the interaction point.
The $\pi^0$ and $\eta$ mesons are reconstructed via $\gamma\gamma$ decays.
For photon candidates, they are also required to pass the criteria for neutral showers in fast simulation.
The $\gamma\gamma$ combinations with invariant masses $M_{\gamma\gamma}\in(0.115,\,0.150)$
and (0.500, 0.570) GeV$/c^{2}$ are regarded as $\pi^0$ and $\eta$ mesons, respectively.
A kinematic fit is performed to constrain $M_{\gamma\gamma}$ to the $\pi^{0}$ or $\eta$
nominal mass.
The $\eta$ candidates for the $\eta\pi^-$
ST channel are also reconstructed
via $\pi^0\pi^+\pi^-$ candidates with its invariant mass
within $(0.530,\,0.570)~\mathrm{GeV}/c^2$.
The $\eta^\prime$ mesons are reconstructed via two decay modes, $\eta\pi^+\pi^-$ and $\gamma\rho^0$,
whose invariant masses are required to be within
$(0.946,\,0.970)$ and $(0.940,\,0.976)~\mathrm{GeV}/c^2$, respectively.
In addition, the minimum energy
of the $\gamma$ from $\eta'\to\gamma\rho^0$ decays must be greater than 0.1 GeV.
The $\rho^0$ and $\rho^+$ mesons are reconstructed from $\pi^+\pi^-$ and $\pi^+\pi^0$
candidates, whose invariant masses are required to be larger than $0.5~\mathrm{GeV}/c^2$
and within $(0.670,\,0.870)~\mathrm{GeV}/c^2$, respectively.
For $\pi^+\pi^-\pi^-$ and $K^-\pi^+\pi^-$ tags, the dominant peaking backgrounds from $D^-_s\to K^0_S\pi^-$ and $D^-_s\to K_S^0K^-$ events with $K_{S}^{0}\to \pi^{+}\pi^{-}$ are rejected by requiring the invariant mass of any $\pi^+\pi^-$ combination to be more than 0.03~GeV/$c^2$ away from the nominal $K^0_S$ mass~\cite{ParticleDataGroup:2020ssz}.

\begin{figure*}[htbp]
	\centering
	\includegraphics[height=12cm, width=18cm]{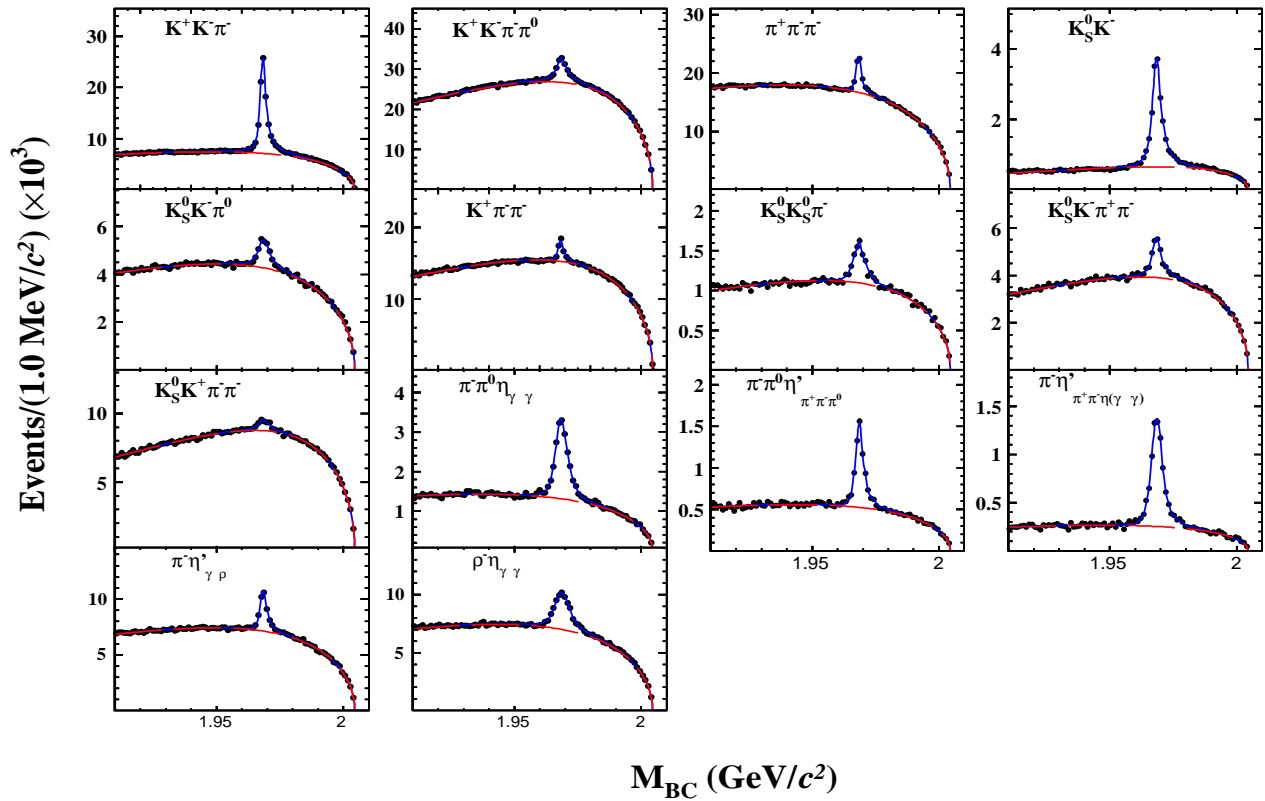}
	\caption{Fits to the $M_{\rm BC}$ distributions of $D_s^-$ candidates. The points with error bars are inclusive MC. The blue curves are the fit results. The red curves are the fitted combinatorial backgrounds.}
	\label{fig:mbc}
\end{figure*}

Two kinematic variables ($\Delta E$, $M_{\rm{BC}}$) reflecting energy and momentum conservation are used to identify the tagged $D_s^-$
candidates. First, we calculate the energy difference
\begin{equation}
	\Delta E = E_{D_s^-} - E_{\rm beam},
\end{equation}
where $E_{D_s^-}$ is the reconstructed energy of a tagged $D_s^-$ meson and $E_{\rm beam}$ is the beam energy. Correctly reconstructed
signal events peak around zero
in the $\Delta E$ distribution. To improve the signal purity,
requirements of $\Delta E$ are applied, corresponding to
$\pm3\sigma_{\Delta E}$ where $\sigma_{\Delta E}$ is the resolution of $\Delta E$ for each tag mode.
If there are multiple $D_s^-$ candidates per tag mode, the one with minimum $|\Delta E|$ is kept for further analysis.
The second variable is the beam-energy constrained mass
\begin{equation}
	M_{\rm BC}=\sqrt{E_{\rm beam}^2/c^4-\overrightarrow{p}_{D_s^-}^2/c^2},
\end{equation}
where $\overrightarrow{p}_{D_s^-}$ is the three-momentum of the tagged $D_s^-$ candidate.
Figure~\ref{fig:mbc}
shows the $M_{\rm{BC}}$ distributions for pseudo-data. The ST yields are obtained by fitting the $M_{\rm{BC}}$
distributions where a MC-determined signal shape is used to model the signal and
an ARGUS \cite{Albrecht:1989ga} function is for background.
To select the signal process with a high purity, a mass window is required on $M_{\rm{BC}}$ within $\pm3\sigma_{M_{\rm{BC}}}$,
where $\sigma_{M_{\rm{BC}}}$ is resolution of $M_{\rm{BC}}$ determined by fitting with a double-Gaussian function.


The $D_s^+ \to \mu^+\nu_\mu $ candidate events are selected in the recoil side of
the tagged $D_s^-$. We require that there is only one candidate charged track in the remaining particles
whose charge is opposite to the tagged $D_s^-$.
The charged track is identified as a muon candidate after passing the corresponding requirements in fast simulation~\cite{fastsimu}.
To suppress the backgrounds with extra photon(s),
the maximum energy of the unused showers ($E_{\mathrm{extra}~\gamma}^{\rm max}$) is required to be less than 0.4 GeV.

The background events survived from above selection criteria can be categorized into two types. The first type, noted as BKGI, contains a correctly reconstructed $D_s^-$ but the signal side is misreconstructed from $D_{s}^{+}\to \tau^{+}\nu_{\tau}$ and other $D_s^+$ decays.
The contribution of BKGI is estimated from exclusive MC samples and the normalized number of events is fixed in the fit.
The second type, noted as BKGII, contains the non-$D_s^+$ background,
which is expected to be a smooth distribution under the $D_s^-$ peak in the $M_{\rm BC}$ spectra. The contribution of BKGII can be estimated from the sideband events of  $M_{\rm BC}$ , defined as (1.915, 1.935) GeV/$c^2$ and (1.990, 2.000)
GeV/$c^2$.

\begin{figure}[htbp]
	
	\centering
	\includegraphics[width=0.5\textwidth]{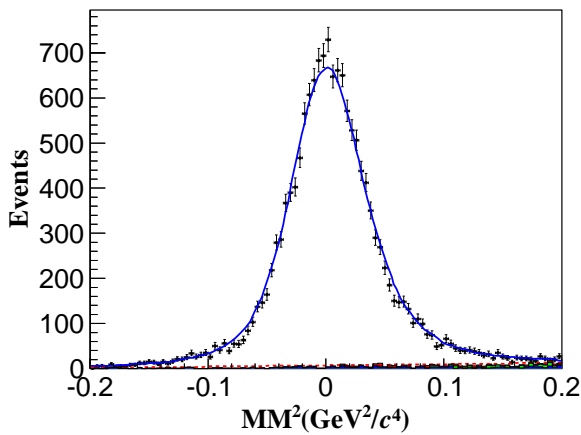}
	
	\caption{ Fit to the MM$^{2}$ distribution of the $D_s^+\to\mu^+\nu_\mu$ candidates.
		Dots with error bars are data. Blue solid curve is the fit result. Red dotted curve
		is the fitted background. Green and blue cross-hatched histograms are the BKGI component, and the
		BKGII component, respectively.}
	\label{fig:m2ss}
\end{figure}

To characterize the signal of $D^+_s\to \mu^+\nu_\mu$, the missing mass squared (MM$^2$) is defined as
\begin{equation}
	\text{MM}^{2}=\left(E_{\rm beam}-E_{\mu^+}\right)^2/c^4-\left(-\overrightarrow{p}_{D_s^-}-\overrightarrow{p}_{\mu^+}\right)^2/c^2
\end{equation}
where $E_{\mu^+}$ and $\overrightarrow{p}_{\mu^+}$ are the energy and momentum of the muon candidate,
respectively.
The signal yield is extracted by fitting the combined MM$^2$ distribution from all $14$ tag modes,
where a shape extracted from signal MC sample is used to describe the signal and a first-order Chebychev polynomial
is used to describe the background, as shown in Fig.~\ref{fig:m2ss}.

The BF of the $D^+_s\to \mu^+\nu_\mu$ is calculated by
\begin{equation}
	\mathcal{B}_{D_s^+ \to \mu^+\nu_\mu}=\frac{N_{\rm sig}}{N_{\rm tag}\times \bar{\epsilon}_{\rm sig}},
	\label{eq:eff}
\end{equation}
where $N_{\rm sig}$ is the number of the signal events determined by a fit to the MM$^2$ spectrum,
and $N_{\rm tag}$ is the number of events for all ST modes by fits to the $M_{\rm BC}$.
The averaged detection efficiency for $D^+_s\to \mu^+\nu_\mu$ can be expressed as
\begin{equation}
	\bar{\epsilon}_{\rm sig} = \Sigma_i\left(\frac{N_{\rm tag}^{i}}{N_{\rm tag}} \times \frac{\epsilon_{\rm tag, \rm sig}^{i}}{\epsilon_{\rm tag}^{i}}\right)
\end{equation}
where $N_{\rm tag}^{i}$ is the number of events for ST mode $i$, $\epsilon_{\rm tag, \rm sig}^{i}$ is the
efficiency of detecting both the ST mode $i$ and the pure leptonic decays, and
$\epsilon_{\rm tag}^{i}$ is the efficiency of detecting the ST mode $i$. The efficiencies of ST modes are 
determined with an independent generic MC sample, and the efficiencies of DT modes are determined with the 
signal MC sample of $e^+e^- \to D_{s}^{+}D_{s}^{-}$, where $D_{s}^{-} \to $ 14
tag modes and $D_s^+\to\mu^+\nu_\mu$.

\begin{figure}[tb]	
	\centering
	\includegraphics[width=0.5\textwidth]{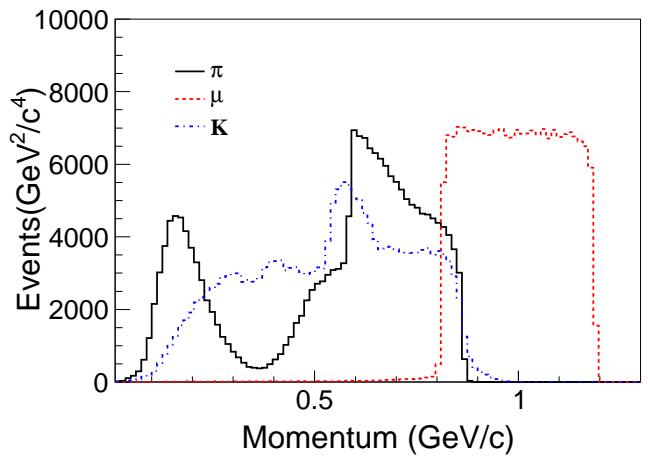}
	\caption{The momentum of $\pi$, $\mu$ and $K$ from $D_s^{-} \to K^{+}K^{-}\pi^{-}, D_s^{+} \to \mu^{+} \nu_{\mu}$ signal MC.}
	\label{mom1}
\end{figure}

\begin{figure*}[htbp]
	
	\centering
	\includegraphics[width=1.0\textwidth]{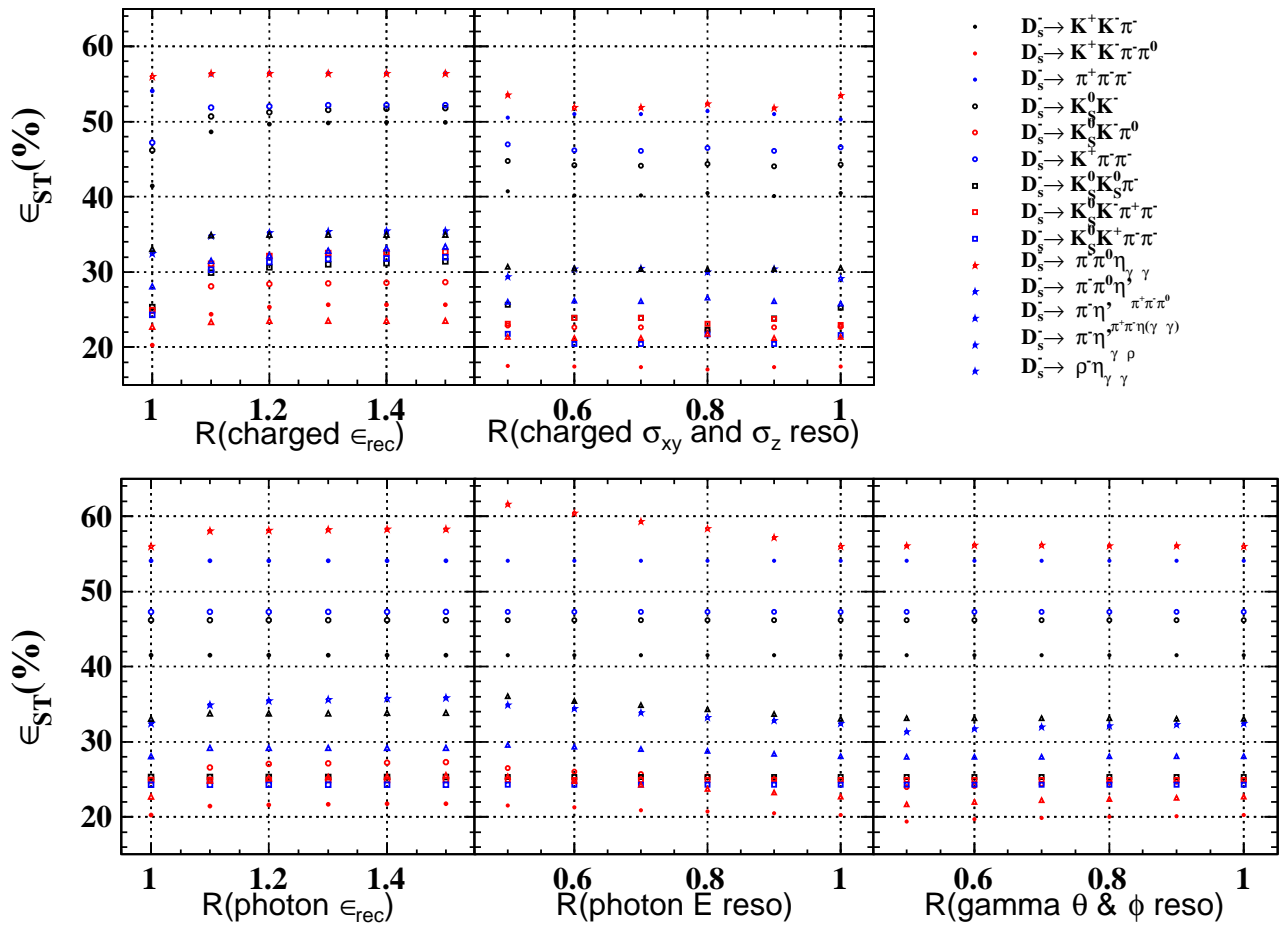}
	
	\caption{The optimization of charged track and photon detection efficiencies and resolutions of 14 tag modes.
		The values of $x$ axis represent the optimized proportion of the initial values of detector of response.}
	\label{fig:trkreff}
\end{figure*}

With 0.1~ab$^{-1}$ pesudo-data,
the number of ST events for $D_s^-$ to 14 decay modes is determined to be $3452605\pm 21268$,
and the number of DT events for $D_s^+\to\mu^+\nu_\mu$ is $14687\pm 142$.
The averaged efficiency $\bar{\epsilon}_{sig}$ of signal process is calculated to be ($75.78\pm 0.07)\%$
by combining 14 tag modes.
The corresponding BF of $D_s^+\to\mu^+\nu_\mu$ is calculated to be $(5.61\pm0.05)\times10^{-3}$.
The uncertainties are statistical only.
The calculated BF agrees well with the input value.
We can easily prospect the statistical sensitivity for the BF of $D_s^+\to\mu^+\nu_\mu$ at STCF with
1~ab$^{-1}$ data as it is proportional to $1/\sqrt{\mathcal{L}}$, to be
$\mathcal{B}_{D_s^+\to\mu^+\nu_\mu}=(5.610\pm0.016)\times10^{-3}$, where the statistical uncertainty is 0.3\%.

Since a full systematic study requires both experimental data and MC,
we are limited in our ability to estimate every possible source.
A more precise estimation of systematic uncertainty will not be feasible until
the design and construction of the detector is completed.
Therefore, a rough systematic uncertainty is estimated by referring
to measurement at BESIII~\cite{Ablikim:2018jun},
which includes reducible systematic uncertainties, $\sigma_{\rm red.}$,
that can be scaled according to luminosity,
and the irreducible systematic uncertainties, $\sigma_{\rm irred.}$, which
mainly comes from theoretical input. It also includes predictably optimized
systematic uncertainties, $\sigma_{\rm pre.}$, which mainly be optimized by control samples and corrections.
The reducible systematic uncertainties include the tracking and PID of $\mu$ detection,
other selection criteria such as extra energy requirement and $D_s^-$ candidate ST by
background function, as well as the uncertainty from the fitting procedure of MM$^{2}$ from background.
The irreducible systematic uncertainties come from the effect of radiative correction.
The predictably optimized systematic uncertainties come \\from fit range, bin size,
signal shape of fitting procedure of $M_{BC}$ and MM$^{2}$, final state radiation~(FSR) and tag bias.
The total systematic uncertainty can be roughly estimated by
$\sigma_{\rm syst.}=\sqrt{\sigma_{\rm red.}^{2}\times(3.19^*\sigma_{4.18})/(1000^*\sigma_{4.01})+\sigma^{2}_{\rm irred.}+\sigma^{2}_{\rm pre.}}$, to be 1.0\%, dominated by irreducible systematic uncertainties.
However, it is expected that theoretical calculation concerning to the radiative $D_s^+ \to \gamma \mu^+ \nu_\mu$ decay will be
able to be improved with better precision.


\section{Optimization of detector response}
\label{sec:OPTIMIZATION}
In the results presented above, a series of optimizations on detector responses have been performed,
including efficiencies of charged
particles and photons,  momentum resolution of charged tracks, energy/position resolution of
photons, and PID efficiencies.
Following we will introduce the details of these optimizations one by one.

~{\it a.~Tracking efficiency of charged particles:} The response of tracking efficiency in fast simulation is characterized by
its transverse momentum $p_{T}$ and polar angle $\cos\theta$.
For high-momentum tracks, {\it e.g.} $p_{T}>0.4$~GeV/$c$ of charged pions,
the tracking efficiency within acceptance is over 99\%.
For low-momentum tracks, {\it e.g.} $p_{T}=0.1$~GeV/$c$ of charged pions,
the tracking efficiency is low due to various effects such as
electromagnetic multiple scattering, electric field leakage, ionization energy loss {\it etc.}.
Figure~\ref{mom1} shows the momentum distribution of charged pions in $D_{s}^{+}$ decay, where there is
a considerable number of particles with momentum lower than 0.4~GeV/$c$.
Benefiting from the flexible interface of changing the response of charged
particles, different tracking efficiencies are set in the simulation
at $p_{T}$ in (0.05, 0.1)~GeV/$c$ of charged pions, where tracking efficiencies at other momentum
are scaled proportionally.
With the tracking efficiency varies from 60.16\% to 90.24\% for $p_{T}$ in (0.05, 0.1)~GeV/$c$,
the detection efficiency of 14 tag modes increases.
The variation of detection efficiency for 14 tag modes
along with different tracking efficiencies
are shown in Fig.~\ref{fig:trkreff},
where the optimized point is found when the tracking efficiency is 72.19\% for charged pions at
$p_{T}$ in (0.05, 0.1)~GeV/$c$.

\begin{figure}[tb]	
	\centering
	\includegraphics[width=0.4\textwidth]{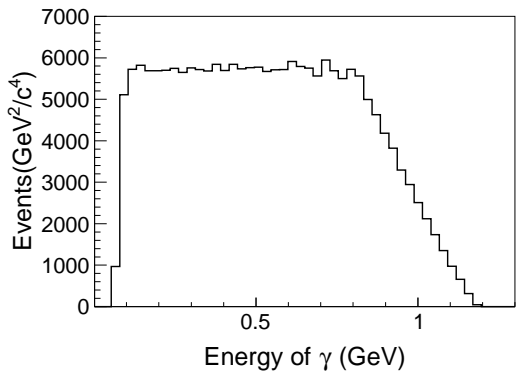}
	\caption{The energy of $\gamma$ from MC truth. }
	\label{mom2}	
\end{figure}
\begin{figure}[tb]	
	\centering
	\includegraphics[width=0.4\textwidth]{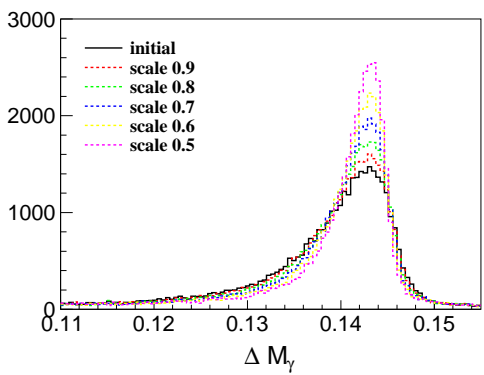}
	\caption{The $\Delta M$ optimization with $\gamma$ energy resolution. }
	\label{fig:deltaMgamma}
\end{figure}

\begin{figure*}[htbp]
	\centering	
	
	\includegraphics[width=0.48\textwidth,height=7cm]{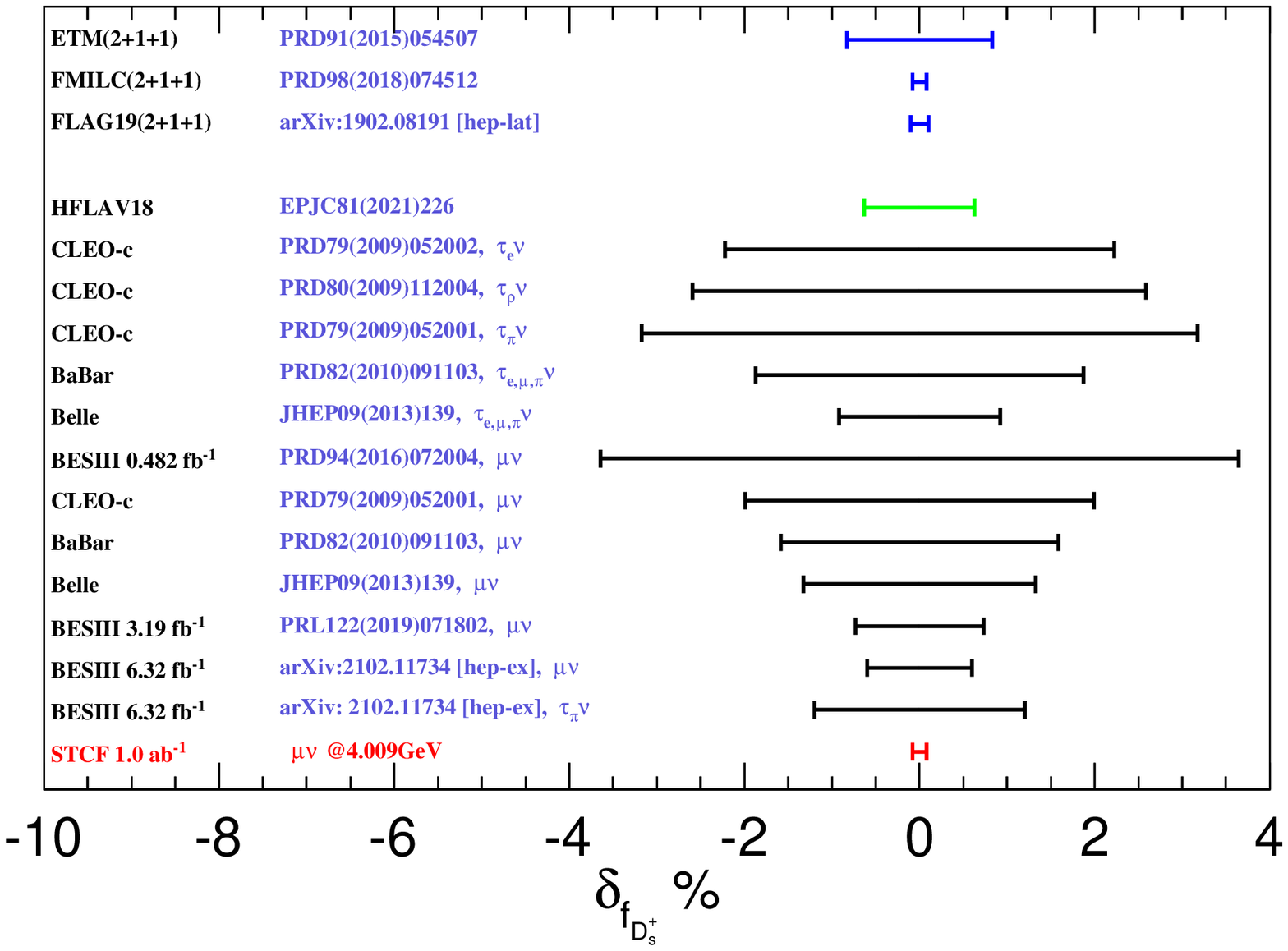}
	\includegraphics[width=0.48\textwidth,height=7cm]{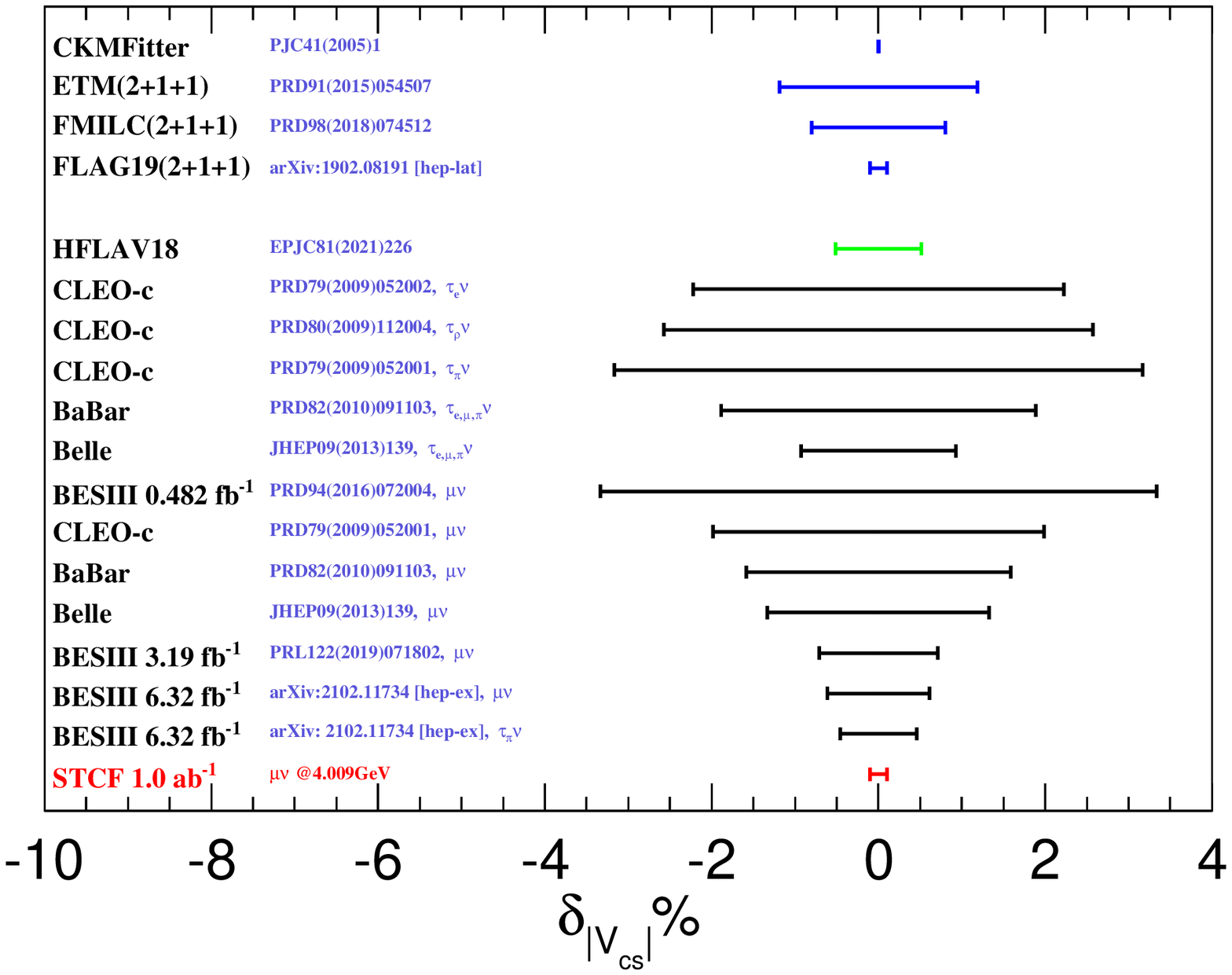}
	\caption{The comparison of $\delta_{f_{D_s^+}}$ (left) and $\delta_{|V_{cs}|}$ (right) with
		previous measurements, the width only represents the statistical uncertainty.}
	\label{fds}
\end{figure*}
\begin{figure}[htbp]

	\includegraphics[width=0.48\textwidth]{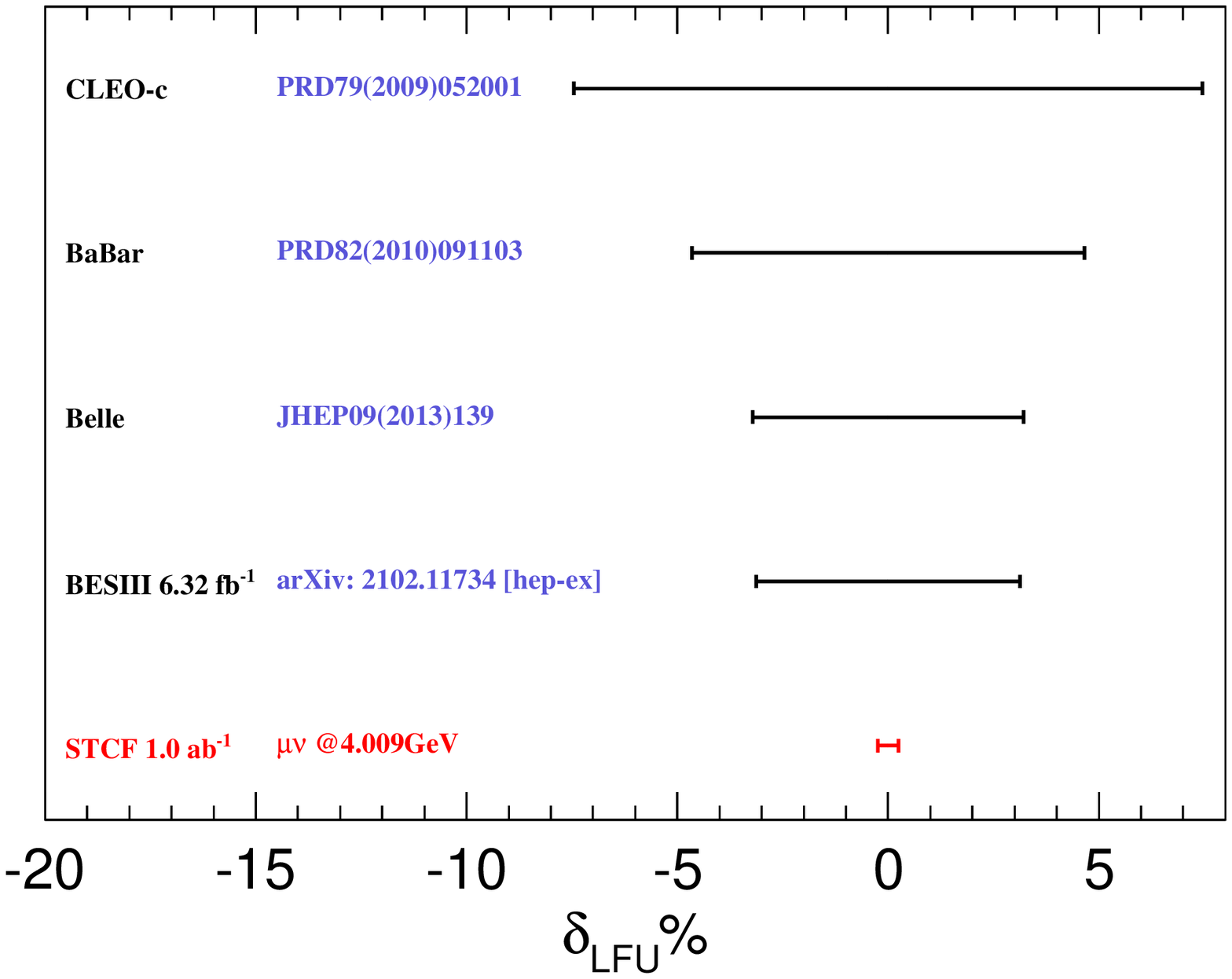}
	\caption{The comparison of $\delta_{LFU}$ with previous measurements, the width only
		represents the statistical uncertainty.}
	\label{LFU}
\end{figure}

~{\it b.~Momentum resolution of charged tracks:}
The momentum resolution of a charged track is affected by the multiple scattering
and the accuracy of the track position from its flight trajectory.
The former one is related with the material, better with lower atomic number, used by the tracker system.
The spatial resolution of the charged track flight trajectory, determined by the accuracy of the
track position, is studied for its influence on momentum resolution by applying a
set of $\sigma_{xy}$ from 65~$\mu$m to 130~$\mu$m
and $\sigma_{z}$ from 1240~$\mu$m to 2480~$\mu$m proportionally.
Since the resolutions of kinematic variable $\Delta E$ and $M_{\rm BC}$ are affected
by the momentum resolutions of charged tracks, different mass windows are then applied
to obtain the detection efficiencies of 14 tag modes.
The results are shown in Fig.~\ref{fig:trkreff} where weak dependence from different spatial
resolution is observed.

~{\it c.~Detection efficiency of photons:}
The energy of photons from $D_{s}^-$ decay ranges from less than 0.1 to 1.2~GeV as shown in Fig.~\ref{mom2}.
The detection efficiency of photons is studied with energy to be 200 MeV.
Figure~\ref{fig:trkreff} shows the detection efficiency of 14 tag modes with the variation
of detection efficiency, where the optimized point is found to be 97.32\% for photons with
energy of 200~MeV.

~{\it d.~Energy/Position resolution of photon:}
The energy and position resolutions of photons are two key parameters for photon detection.
Various resolutions of photons, {\it i.e.} energy resolution from 2.5\% to 1.25\% and position
resolution from 6~mm to 3~mm at 1~GeV, have been tested for 14 tag modes of $D_{s}^-$ decay.
It is found that the detection efficiencies of decay modes containing photons
are improved, especially for better energy resolution, as shown in Fig.~\ref{fig:trkreff}.

The influence of energy and position resolutions of photons are also studied with the process
$e^{+}e^{-}\to D_{s}^{+}D_{s}^{*-}+c.c$, where the energy of photon in $D_{s}^{*-}\to \gamma/\pi^{0} D_{s}^{-}$
locates within 200~MeV. Two key parameters of this process are discussed. One is the mass difference
of reconstructed $D_{s}^{*-}$ and $D_{s}^{-}$, $\Delta M$, whose
resolution is determined by resolution of photon energy.
The other one is the $\pi^{0}/\gamma$ contamination in reconstruction
of $D_{s}^{*-}\to \gamma/\pi^{0} D_{s}^{-}$.
After a series of test with different energy/position resolutions, it is found that
a better energy resolution will improve the resolution of $\Delta M$ as shown in Fig.~\ref{fig:deltaMgamma}.
No obvious improvement
is observed  for the resolution of $\Delta M$ in the variation of position resolution,
and the $\pi^{0}/\gamma$ contamination rate in different energy or position resolutions
of photon keeps unchanged, to be about 28.1\%. Although the volume of $\pi^{0}$ from
$D_{s}^{*-}$ is pretty small, it does make an impact.

{\it e. $\pi/K$ identification:}
The identifications of $\pi$ and $K$ are essential for the charm physics at STCF.
Since the momenta of $\pi/K$ are relatively low in this analysis, they can be mostly
identified by the characteristic ionization energy loss~($dE/dx$) in the tracker system.
Simulation indicates that with a $dE/dx$ resolution of 6\%, the $\pi/K$ can be well
separated when $p<0.8$~GeV/$c$, which can meet the requirement for $\pi/K$ separation in this analysis.

{\it f. $\mu$ identification: }
The momentum of $\mu$ in this analysis is shown in Fig.~\ref{mom1}, where the MUD is expected to provide
a high identification efficiency for muon and low $\pi/\mu$ mis-identification rate.
Reference~\cite{zhujun} describes the details of the baseline design of MUD at STCF. With the performance of MUD
provided in Ref.~\cite{zhujun},  three $\pi/\mu$ mis-identification rates are tested, to be 1\%, 1.6\% and 3\%,
corresponding to the identification efficiencies of muon to be 85\%, 92\% and 97\% at $p_{\mu}=1$~GeV/$c$.
The optimized result is achieved at 3\% of $\pi/\mu$ mis-identification rate.

From the discussion above, we use the following detector responses optimized from this analysis while others
are kept the same in the fast simulation.
The optimized responses include a tracking efficiency for low momentum charged particles, to be 72.19\% at
$p_{T}=0.1$~GeV/$c$,
a detection efficiency of photons, to be 97.32\% at 0.2 GeV,
a $\pi/\mu$ mis-identification rate of 3\% with the MUD performance provided from Ref.~\cite{zhujun}.
Comparing to the default response provided by fast simulation,
the detection efficiency for ST is increased by a factor that in the range between 1.1 and 1.2, depending on the tag modes,
and the efficiency for selecting $D_{s}^{+}\to\mu^{+}\nu_{\nu}$ is increased by a factor of 1.3.

\section{Results and Discussion}
\label{results}
With the expect sensitivity of ${\mathcal{B}}_{D_s^+\to\mu^+\nu_\mu}= (5.610\pm0.016)\times10^{-3}$ at STCF obtained in this analysis, and the world average values of  $G_F$, $m_\mu$, $m_{D^+_s}$ and the lifetime of $D_s^+$~\cite{ParticleDataGroup:2020ssz} as listed in Table~\ref{parameters},
the product value of $f_{D_s^+}|V_{cs}|$ can be obtained according to Eq.~(\ref{eq1})
\begin{equation}
	f_{D_s^+}|V_{cs}|=248.9\pm0.4_{\rm stat.}~\mathrm{MeV}. \nonumber
\end{equation}
Taking the CKM matrix element $|V_{cs}|=0.97320\pm{0.00011}$ from the global fit
in the SM~\cite{ParticleDataGroup:2020ssz} or the averaged decay constant
$f_{D_s^+}=249.9\pm0.5~\mathrm{MeV}$ of recent LQCD calculations~\cite{Bazavov:2014wgs,Carrasco:2014poa}
as input, $f_{D_{s}^{+}}$ and $|V_{cs}|$ can be determined separately
\begin{equation}
	f_{D_s^+}=255.8\pm0.4_{\rm stat.}~\mathrm{MeV} \nonumber
	\label{eq8}
\end{equation}
and
\begin{equation}
	|V_{cs}|=0.996\pm0.002_{\rm stat.}. \nonumber
	\label{eq9}
\end{equation}

\begin{table}[htbp]
	\centering
	\caption{The values of parameters taken from PDG~\cite{ParticleDataGroup:2020ssz}}
	\label{parameters}
	\begin{tabular}{c|c}	
		\hline\hline
		~~~Parameters~~~ & ~~~Value~~~~  \\
		\hline
		$G_{F}$    & $(1.1663787 \pm 0.0000006)\times10^{-5}$ GeV$^{2}$   \\
		$m_{\mu}$  & $105.6583745 \pm 0.0000024 $  MeV/$c^{2}$  \\
		$m_{D^{+}_{s}}$ & $1968.35 \pm 0.07 $ MeV/$c^{2}$\\
		$\tau_{D_{s}^{+}}$  & $(504 \pm 4)\times10^{-15}$s \\
		\hline
	\end{tabular}
\end{table}

Besides, the LFU can be tested with the ${\mathcal{B}}_{D_s^+\to\mu^+\nu_\mu}$ obtained in this study
combining the $\mathcal{B}_{D_s^+\to\tau^+\nu_\tau}=(5.49\pm0.02)\%$ from STCF,
to be
\begin{equation}
	\frac{\mathcal{B}_{D_s^+\to\tau^+\nu_\tau}}{\mathcal{B}_{D_s^+\to\mu^+\nu_\mu}}=9.79\pm0.05. \nonumber
\end{equation}
It is worth mentioning that the $f_{D_s^+}$ and $|V_{cs}|$ are from recent LQCD calculations,
whose uncertainty will be reduced in the future, therefore the
$f_{D_s^+}$ and $|V_{cs}|$ determined at STCF will also be reduced in the future. As shown in
Figs.~\ref{fds} and \ref{LFU}, the uncertainty of $f_{D_s^+}$ we determined
is smaller than that from LQCD calculation, which calls for further improved LQCD calculation.
The expected uncertainty of $|V_{cs}|$ is close to that from
LQCD calculation. The accuracy of LFU test can be improved obviously experimentally,
which makes it promising to search for the new physics beyond the SM.

\section{Acknowledgments}

The authors are grateful to the software group of STCF and the physics group of STCF for the profitable discussions.
We express our gratitude to the supercomputing center of USTC and Hefei Comprehensive
National Science Center for their strong support.
This work is supported by the Double First-Class university  project foundation of USTC, the Natural Science Foundation of Hunan Province under Contract Nos. 2020RC3054 and 2019JJ30019, and the National Natural Science Foundation of China under Projects No. 11625523.

\bibliography{ljjbib}

\end{document}